\newif\ifonecol % Never comment out this line
\onecolfalse % Two-columns

\ifonecol
\documentclass[journal,10pt,onecolumn]{IEEEtran}
\else
\documentclass[journal]{IEEEtran}
\fi

\IEEEoverridecommandlockouts
\usepackage{cite}
\usepackage{amsmath,amssymb,amsfonts,bbm}
\usepackage{algorithmic}
\usepackage{graphicx}
\usepackage{textcomp}

\usepackage{setspace}
\usepackage{latexsym}
\usepackage{graphicx}
\usepackage{comment}
\usepackage{hyperref}
\usepackage[dvipsnames]{xcolor}
\newcommand{\RR}{\mathbb{R}} % real set
\newcommand{\ee}{{\rm e}}
\newcommand{\jj}{{\rm j}}  % imaginary unit
\newcommand{\dd}{{\rm\,d}} % differential (for integrals)

\newcommand{\av}{{\bf a}}

\newcommand{\gv}{{\bf g}}

\newcommand{\pv}{{\bf p}}

\newcommand{\uv}{{\bf u}}

\newcommand{\vv}{{\bf v}}

\newcommand{\Am}{{\bf A}}

\newcommand{\Rm}{{\bf R}}

\newcommand{\Um}{{\bf U}}

\newcommand{\Vm}{{\bf V}}

\newcommand{\Cc}{{\cal C}}

\newcommand{\Fc}{{\cal F}}
\newcommand{\Gc}{{\cal G}}

\newcommand{\Ic}{{\cal I}}

\newcommand{\Lc}{{\cal L}}

\newcommand{\Pc}{{\cal P}}

\newcommand{\Tc}{{\cal T}}
\newcommand{\Zc}{{\cal Z}}

\def\trace{\mathsf{Tr}}

\def\Tran{^{\mathsf{T}}}

\def\ben{\begin{enumerate}}
\def\beq{\begin{equation}}
\def\beqa{\begin{eqnarray}}
\def\bit{\begin{itemize}}
\def\een{\end{enumerate}}
\def\eeq{\end{equation}}
\def\eeqa{\end{eqnarray}}
\def\eit{\end{itemize}}

\def\non{\nonumber\\}

\begin{document}

\title{Spectrum-Aware IRS Configuration Techniques for Ultrawideband Signals}

\author{Alessandro~Nordio,~\IEEEmembership{Member, IEEE}, Alberto~Tarable,~\IEEEmembership{Member, IEEE}, Francisco~J.~Escribano,~\IEEEmembership{Senior Member, IEEE}
%}

\thanks{A. Nordio and A. Tarable are with the National Research Council of Italy, Institute of Electronics, Information Engineering and Telecommunication  (CNR-IEIIT), 10129 Torino, Italy (e-mail: alessandro.nordio@cnr.it, alberto.tarable@cnr.it).}
\thanks{F. J. Escribano is with the Signal Theory and Communications Department at Universidad de Alcalá, 28805 Alcalá de Henares, Spain (email: francisco.escribano@uah.es).}
}

\maketitle

\begin{abstract}
Intelligent reflecting surfaces (IRS) have become the subject of many current research efforts, as the ongoing wireless spectrum crunch has made the need to open higher frequency bands a priority. IRS are one of the alternatives proposed to overcome the problem of line-of-sight blocking in very high frequency wireless scenarios. The current state-of-the-art shows the difficulty of implementing practical IRS designs able to redirect large signal bandwidths, prone to the so-called beam split (BS) dispersion effect. In this work, we propose two highly efficient configuration techniques, adapted to ultrawideband downlink scenarios, based on localized optimization over the IRS surface. Such techniques exploit the BS effect while taking into account for the shape of the transmitted signal spectrum. Simulations considering different geometrical setups and different signal spectra show how the proposed techniques are able to guarantee an increased signal power at the receiver with respect to classical narrowband-based solutions or techniques that perform a global optimization over the entire IRS surface.
\end{abstract}


%\section{Single antenna BS and UE, linear IRS, narrowband signal}
%Let the position of the BS and UE be $(x_{\rm BS},y_{\rm BS})$ and $(x_{\rm UE},y_{\rm UE})$
%The IRS is located on the $x$ axis.
%
%The signal scattered by the IRS element at position $x$ and recevived at the UE is proportional to
%\[ y(t,x) \propto s(t) \ee^{-\jj 2 \pi \frac{d_1(x)+d_2(x)}{\lambda}} \ee^{\jj \phi(x)} \frac{1}{d_1(x) d_2(x)} \]
%where $s(t)$ is the (monochromatic) transmitted signal
%\[  d_1(x) = \sqrt{(x-x_{\rm BS})^2+y_{\rm BS}^2}, \qquad d_2(x) = \sqrt{(x-x_{\rm UE})^2+y_{\rm UE}^2} \]
%and $\phi(x)$ is the phase shift imposed by the IRS at position $x$.
%In order to align the signal phase at the UE we need to impose
%\[ \phi(x) = \frac{2 \pi}{\lambda} (d_1(x)+d_2(x)) = \frac{2\pi}{\lambda}\left(\sqrt{(x-x_{\rm BS})^2+y_{\rm BS}^2}+\sqrt{(x-x_{\rm UE})^2+y_{\rm UE}^2}\right) \]
%  and the phase-gradient is
%\begin{eqnarray}
%  \frac{\dd \phi(x)}{\dd x}
%  &=&\frac{2\pi}{\lambda}\left(\frac{x-x_{\rm BS}}{\sqrt{(x-x_{\rm BS})^2+y_{\rm BS}^2}}+\frac{x-x_{\rm UE}}{\sqrt{(x-x_{\rm UE})^2+y_{\rm UE}^2}}\right) \non
%  &=&\frac{2\pi}{\lambda}\left(\frac{\frac{x-x_{\rm BS}}{y_{\rm BS}}}{\sqrt{1+\frac{(x-x_{\rm BS})^2}{y_{\rm BS}^2}}}+\frac{\frac{x-x_{\rm UE}}{y_{\rm UE}}}{\sqrt{1+\frac{(x-x_{\rm UE})^2}{y_{\rm UE}^2}}}\right) \non
%  &=&-\frac{2\pi}{\lambda}\left(\sin \psi_i(x)+\sin \psi_r(x)\right)            
%\end{eqnarray}
%which is consistent with the generalized Snell law. ($\psi_i(x)$ and  $\psi_r(x)$ are the BS and UE angles as observed from the IRS.

\section{Introduction}

\IEEEPARstart{T}{Hz} communications and reflective intelligent
surfaces are expected to be key technologies characterizing the future
sixth generation (6G) of mobile communications. THz communications
allows networks to operate in the 0.1--10\,THz frequency range and
reach data rates of terabits per second~\cite{Saqib23}. With such technology, wireless devices will
be able to transmit ultra-wideband (UWB) signals, up to several hundred
GHz, representing a significant fraction (20\% or more) of the central
frequency.

However, to obtain such performance, there are several challenges that
need to be faced, mainly due to the harsh propagation environment at
sub-millimeter wavelengths, which includes severe path losses and
blockages, even by thin obstacles. These issues can be addressed by
using large high-gain antenna arrays and beamforming techniques, as
well as deploying relays or intelligent reflective surfaces
(IRSs)~\cite{DeLima21,Naeem22}. In particular, IRSs are passive metasurfaces
composed of a large number of sub-wavelength size elements that are
able to apply adjustable phase shifts to the impinging
signal. Macroscopically, they behave as anomalous mirrors, according
to the generalized Snell's law~\cite{GeneralizedSnell}, able to
reflect the signal towards a desired direction. Thus, they are
effective in assisting communication between pairs of wireless network nodes,
especially when a direct line-of-sight (LoS) link among them is not
available~\cite{Wu21}.

Another challenge arises from the  interaction between UWB signals and large antenna arrays, as well as IRSs. 
Transceivers equipped with large uniform linear or planar antenna arrays (ULAs/UPAs) are able to generate narrow beams towards a desired direction, thus concentrating the electromagnetic power only where is needed and reducing interference. 
However, when the array has many antenna elements, the difference in signal propagation delay between different elements can become a significant portion of the symbol period. 
This phenomenon, known in the array and radar signal processing literature as the {\em spatial-wideband effect}~\cite{SpatialWideband}, can lead to performance losses. 
In the frequency domain, it is often referred to as {\em beam squint} or {\em beam split} (BS)~\cite{Delay_phase_precoding2019,WB_beamforming_mMIMO}. 
The term ``beam squint'' is typically used in the context of mmWave communications, whereas ``BS'' is preferred in the THz band, which can accommodate signals with much larger bandwidth.
Because of this phenomenon, beams generated by ULAs/UPAs or IRSs do not point in the same direction at different frequencies. 
Consequently, only certain frequencies benefit from high array gain towards the desired direction, while the energy associated to others may not be efficiently used, and in the worst case is spread in the environment and wasted. 
This effect becomes more significant as the size of the antenna array and the signal bandwidth increase.
As an example, the angular deviation of a beam due to the BS effect is about $6^\circ$ for a $10\%$ relative
bandwidth and can be significantly larger for UWB signals having relative bandwidth in excess of $20\%$.
Without countermeasures, an IRS-aided communication channel, like the one depicted in Fig.~\ref{fig:irs_3D}, is susceptible to BS effects caused by both the transmitter antenna array and the IRS.

Moreover, in a scenario where large IRSs or antenna arrays are employed and
where the wireless links are short, as in indoor communications, the
receiver may be located in the radiative near-field region of the
transmitter. In such conditions, which require to consider spherical
wavefronts instead of plane waves, the design of communication systems
is even more challenging, since beams are not only characterized by their
direction, as in far field, but also by their distance to the point in space which they are focused to.

In summary, the design of near-field communication systems that involve ultra-wideband signals, large antenna arrays, and IRSs poses the following challenges, which must be considered to improve efficiency
and performance~\cite{Droulias24}:
\begin{itemize}

  \item handling of the BS effect generated by the transmit antenna arrays, in order to increase the amount of signal energy transferred to the destination, with consequent improved performance;

  \item design of IRSs whose operational bandwidth is large enough to
  support UWB signals;

  \item efficient configuration of the phase shifts of the IRS elements, so as to compensate the BS generated by both the transmit antenna arrays and the IRS itself. Solutions must be aware of the system geometry and of the spectral characteristics of the signals.
\end{itemize}

\subsection{State of the art}
We here summarize the state of the art of the research regarding the above mentioned challenges.

BS mitigation in antenna arrays is addressed in numerous
studies and resolved using methods such as delay-phase precoding
techniques~\cite{Delay_phase_precoding2022} or true-time delayers
(TTDs) to achieve UWB beamforming. The study
in~\cite{Hashemi2008} explores the implementation of TTDs and their
applications for signals with instantaneous ultra-wide bandwidths. 
To mitigate the loss of gain from the transmit antenna array,~\cite{Chen2025} proposes a new wideband beam alignment framework based on TTDs, which can control BS.
In contrast, \cite{WB_beamforming_mMIMO} considers approaches for eliminating BS based on
virtual sub-arrays and TTD lines; this architecture can nearly match the performance of fully digital transceivers, although it demands high hardware costs and significant power consumption.  
As an alternative, a hybrid solution called beam-split-aware beamforming was recently introduced in~\cite{Elbir_2023} to effectively compensate the impact of BS due to the use of analog beamformers. 
This solution makes use of a single analog beamformer supported by digital beamformers and time-delayers.
Overall, the above-mentioned techniques add a layer of complexity to the transmitter making it more expensive and more difficult to manage. 
Note that, if the IRS is very large compared to its distance from the transmitting array (see Fig.~\ref{fig:irs_3D}), it is not necessary to completely eliminate BS at the transmitter, as long as beams that deviate from their nominal direction eventually impinge on the IRS.

The above proposed techniques are, unfortunately, not suitable for eliminating BS at IRSs. This is because IRSs consist of passive elements with phase shifts that cannot be controlled in the frequency domain.
Some specific solutions have been proposed to reduce the effect of BS at the IRS. 
For example,~\cite{Yashvanth2025} mitigates the spatial-wideband effect and the resulting BS by coordinating multiple cooperating IRSs. 
The spatial diversity provided by BS is instead exploited in an Orthogonal Frequency-Division Multiple Access (OFDMA) communication system~\cite{Siddhartha2025}, by opportunistically scheduling different UEs on different subcarriers. Such game-shifting idea has the weakness of relying on a scarcely controllable phenomenon (BS) as a system resource. 
The work in~\cite{DelayAdjustable_Hanzo} introduces an IRS implementation called delay-adjustable metasurface (DAM), where elements use varactor diodes to impose a controllable extra delay on the reflected signals. A similar approach is taken in~\cite{Sun22}, where small portions of the IRS are connected to a time-delay unit. This approach is also followed in the very recent work~\cite{Qiu25}, where, in a multiuser scenario, the delays and the phase shifts of an IRS are jointly optimized with the beamformer, to maximize the weighted sum rate.
However, DAMs are still in the early concept stage and are subject to power losses, unlike standard IRSs, whose prototypes are already available, as reported in~\cite{prototype}.

Many studies model the response of IRS elements as
frequency-independent in both phase and amplitude, even though this is not accurate in practice~\cite{survey}. This assumption holds reasonably well only within the IRS's {\em operational bandwidth},
which in most applications represents a small fraction of the carrier frequency. Consequently, IRSs have not been
considered suitable yet for supporting UWB signals, which typically have
relative bandwidths exceeding $20\%$. However, a recent advancement in
IRS design introduced a model functioning at
$26.5\,$GHz~\cite{Behrooz2023}, exhibiting an operational bandwidth
around $30\%$ of the carrier frequency, thus paving the way for IRS-aided UWB
communications. As technology continues to evolve, we think that the
capability of handling such large bandwidths will extend to much higher frequencies, reaching the sub-THz range.

The challenge of optimally configuring an IRS for wideband signals has only recently been considered. The work in \cite{BeamSquintMitigating} used Orthogonal Frequency-Division Multiplexing (OFDM) to maximize the achievable rate, but with a narrow relative bandwidth (about 7\% at 28~GHz) compared to this work. Their method relies on singular-value decomposition applied only to the IRS-receiver link, neglecting the transmitter-IRS channel and the signal's power spectral density.
 
In ~\cite{Tarable2025, NoiICC}, a similar scenario is considered, and some solutions for IRS configuration are presented, among which one is based on the singular-value decomposition of the entire channel matrix between the transmitter and the receiver. 
However,~\cite{Tarable2025, NoiICC} mostly concentrate on a system where the BS at the transmitter is not present, unlike the present work, which instead takes advantage of this phenomenon to realize a spectrum-aware and spatially frequency-selective IRS configuration, with a good trade-off between performance and complexity.

\subsection{Contributions}

%However, in this work, we propose configuring
%the IRS elements appropriately to compensate for the beam-split effect
%in the presence of UWB signals.
This paper considers a non-line-of-sight point-to-point link between an access point (AP) and a user equipment (UE), in the THz band, where an IRS is providing the connectivity thanks to smart reflection. Our main goal is to give simple, yet efficient ways of configuring the IRS so as to maximize the signal power received at the UE, while taking into account the spectrum of the transmitted signal. The main contributions of this paper are the following.

\begin{itemize}
    \item We give a flexible and complete model of the signal received at the UE from the AP, after reflection on the IRS. Unlike other papers, the model distinguishes between the controllable and the uncontrollable parts of the IRS response.  
    \item We formulate the problem of the IRS optimization in the wideband regime, by considering performance figures that take into account the shape of the transmitted signal spectrum.
    \item We derive an upper bound to the modulus of the channel transfer function under the hypothesis that the IRS phase shifts are arbitrary functions of frequency.
    \item In the practical case in which IRS phase shifts are constant with frequency, we propose a spatially frequency-selective approach, in which the IRS phase shift performed by a given element is tuned to a specific position-dependent frequency component of the transmitted signal.
    \item We present a series of numerical results, in realistic geometric scenarios, that show that our proposed approach yields superior performance with respect to narrowband-based methods, and comparable performance with respect to more complex techniques based on eigenvalue decomposition. 
\end{itemize}

Overall, our paper clarifies the achievable performance by using an IRS in a wideband scenario, and our proposed approach yields an effective, relatively low-complexity method to configure the IRS for a transmitted signal with an arbitrary spectrum. Differently from previous works, our proposed solution capitalizes on the BS at the transmitter and the IRS, instead of eliminating it, thus avoiding costly and/or futuristic hardware solutions, and achieving, so to speak,  software-defined IRS-aided UWB far- and near-field communications.

\subsection{Mathematical notation}
Boldface uppercase and lowercase letters denote matrices and vectors,
respectively. The transpose of matrix $\Am$ is denoted by $\Am\Tran$,
whereas $[\Am]_{i,j}$ indicates its $(i,j)$-th element. Coordinate
systems and sets are denoted by calligraphic capital letters. The
$\ell^2$--norm of a vector $\vv$ is denoted by $|\vv|$.

Coordinate systems can be either Cartesian or spherical. The
Cartesian coordinates are denoted by the letters $x,y$, and $z$,
whereas for spherical coordinates we use the letters $\rho$, $\theta$
and $\psi$ to denote radial distance, zenith angle and azimuthal
angle, respectively. The rules for converting Cartesian to spherical
coordinates can be found, e.g. in~\cite{SphericalCoordinates}.

A point $P$ in the Cartesian system $\Cc$ has coordinates
$\pv_{\Cc}=(x,y,z)\Tran$. The same point has coordinates
$\pv_{\Cc'}=(x',y',z')\Tran$ in the Cartesian system $\Cc'$.

\subsection{Paper organization}
The rest of the paper is structured as follows.
Section~\ref{sec:model} introduces the geometric model of the communication system, whereas Section~\ref{sec:signal} specifies the assumptions on the channel and on the transmitted signal, and fully characterizes the transfer function of the complete channel between the AP and the UE. 
Moreover, Section~\ref{sec:signal} formalizes the optimization problem of the IRS configuration. 
Section~\ref{sec:upper_bound} gives a useful upper bound on the magnitude of the channel transfer function. We review in Section~\ref{sec:baseline} some baseline IRS configuration options, which serve as benchmarks for our proposals. In Section~\ref{sec:local}, we present our local optimization techniques, which leverage on signal spectrum, BS at the AP and IRS size to achieve spatial frequency selectivity. In Section~\ref{sec:results}, we show simulation results in different settings, highlighting the advantages and shortcomings of the different IRS optimization techniques. Finally, in Section~\ref{sec:conclusions}, we draw some conclusions.

\section{Geometric model\label{sec:model}}
\begin{figure}[t]
\centerline{\includegraphics[width=0.9\columnwidth]{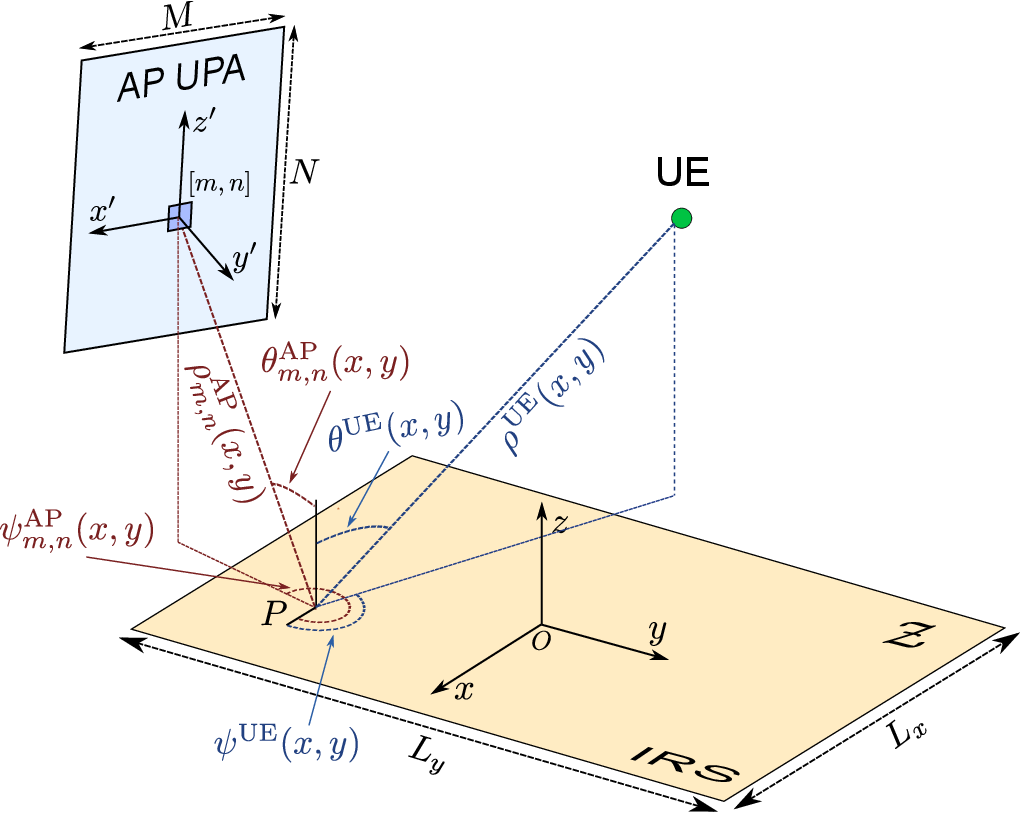}}
\caption{Network geometry.}
\label{fig:irs_3D}
\end{figure}

We consider the downlink of a UWB sub-THz
communication system where an AP communicates with a
UE in an IRS-assisted smart radio environment, as
depicted in Figure~\ref{fig:irs_3D}. The system operates at central
frequency $f_0$.

The IRS lies on the $x-y$ plane of a Cartesian global coordinate
system (GCS) $\Gc$, whose origin coincides with the IRS center, as
depicted in Figure~\ref{fig:irs_3D} \!. The $z$ axis of the GCS is thus orthogonal to the IRS surface, which has rectangular shape, size $L_x \times L_y$, and is made of square-shaped elements separated by distance $\Delta=\lambda_0/2$, where $\lambda_0=c/f_0$, $c$ being the speed of light. The IRS surface is defined in $\Gc$ by the set \[ \Zc = \left\{ (x,y,z)\in \RR^3 \Big| z{=}0, -\frac{L_x}{2}{\le} x {\le} \frac{L_x}{2},
-\frac{L_y}{2}{\le} y {\le} \frac{L_y}{2}\right\}.\]

% As a first approach
%for the analytic work, the IRS surface and its effect on the impinging
%EM waves will be treated as continuous, with $\mathcal{Z}$ denoting
%the set of coordinates of the IRS. In order to refer to a point over
%the IRS surface w.r.t. the GCS, whenever there is no ambiguity, we
%would in general just refer to coordinates $x$ and $y$.

The AP is equipped with a uniform planar array (UPA) of antennas
composed of $M \times N$ elements, spaced by $\Delta$. To the AP UPA
a local coordinate system (LCS) $\Lc$ is associated, as shown in
Figure~\ref{fig:irs_3D} \!, where the UPA lies on its $x'-z'$ plane. The
center of the AP UPA coincides with the origin of $\Lc$. Furthermore,
the orientation of $\Lc$ with respect to $\Gc$ is characterized by
bearing, downtilt and slant angles denoted as $\alpha$, $\beta$, and $\gamma$,
respectively. The origin of $\Lc$ has Cartesian coordinates
$\av_{\Gc}=(x^{\rm AP}, y^{\rm AP}, z^{\rm AP})\Tran$ in $\Gc$. The antenna element of the AP UPA at position $(m,n)$, $m=0,\ldots, M-1$,
$n=0,\ldots, N-1$, is denoted by the point $A_{m,n}$, which has Cartesian coordinates $(\av_{m,n})_{\Lc}=
((n-(N-1)/2)\Delta,0, (m-(M-1)/2)\Delta)\Tran$ in $\Lc$ and
$(\av_{m,n})_{\Gc}$ in $\Gc$. The UE is equipped with a single
antenna and its location is $\uv_\Gc=(x^{\rm UE}, y^{\rm UE}, z^{\rm
UE})\Tran$ in $\Gc$.

Referring to Fig.~\ref{fig:irs_3D}, consider a generic point $P$ on
the IRS surface, with coordinates $\pv_{\Gc}=(x,y,0)\Tran$. We denote
by $\rho^{\rm AP}_{m,n}\left(x,y\right)$ the distance between the
element $(m,n)$ of the AP ULA and $P$.  In a spherical coordinate
system with origin at $P$ and whose axes are oriented as the axes in $\Gc$, the
point $A_{m,n}$ on the AP UPA has coordinates $(\rho^{\rm
AP}_{m,n}(x,y), \theta^{\rm AP}_{m,n}(x,y), \psi^{\rm
AP}_{m,n}(x,y))\Tran$. Similarly, in the same coordinate system, the
UE location has coordinates $(\rho^{\rm
UE}(x,y), \theta^{\rm UE}(x,y), \psi^{\rm UE}(x,y))\Tran$ when
observed from $P$.

The distance $\rho^{\rm AP}_{m,n}$ can be readily obtained as 
\begin{eqnarray}
\label{eq:d1mn}
\rho^{\rm AP}_{m,n}\left(x,y\right) &=& \left|\pv_{\Gc}-(\av_{m,n})_{\Gc}\right| \non
&=&\left|\pv_{\Gc}-\av_{\Gc} - \Rm (\av_{m,n})_{\Lc}\right|
\end{eqnarray}
where we recall that $(\av_{m,n})_{\Gc}$ denotes the coordinates of
the AP UPA element $(m,n)$ in $\Gc$ and $\Rm$ is a $3\times 3$
rotation matrix converting the components of $\av_{m,n}$ from the
coordinate system $\Lc$ to $\Gc$.  The matrix $\Rm$ is function of the
angles $\alpha,\beta$, and $\gamma$. Its expression can be found e.g. in~\cite{3gppchanmodel}.
%{\color{red} Do
%we really need to prove this? The proof of~\eqref{eq:d1mn} is reported
%in Appendix~\ref{app:d1mn}.}

Similarly, we denote by $\rho^{\rm
UE}(x,y)$ the distance between $P$ and the UE, given by
\begin{equation}
\label{eq:d2mn}
\rho^{\rm UE}\left(x,y\right) = \left|\pv_{\Gc}{-}\uv_{\Gc}\right|=\sqrt{(x{-}x^{\rm UE})^2+(y{-}y^{\rm UE})^2+(z^{\rm UE})^2}
\end{equation}

\section{Channel and signal model\label{sec:signal}}
We assume that the LoS link connecting the AP and the
UE is obstructed. However, the connectivity is granted by the IRS, which
reflects the AP signal towards the UE. The transmitted signal
has bandwidth $B$ around central frequency $f_0$, and its spectrum has support $\Fc
\subseteq \Ic_B \triangleq \left[ f_0-\frac{B}{2}, f_0+\frac{B}{2}\right]$. We assume that $B$ is a
substantial fraction of $f_0$, as typical for UWB communications.

Considering the AP--IRS link, the signature of the AP UPA at frequency $f \in \Fc$ as observed from $P\in \Zc$ can be denoted by the $M\times N$ matrix $\Um(x,y,f)$. Its $(m,n)$-th element is given by (\cite{NoiTWC2022})
\begin{equation}
\label{eq:BSsignature}
\left[\Um(x,y,f)\right]_{m,n} = \frac{1}{\sqrt{4\pi} \rho^{\rm AP}_{m,n}\left(x,y\right)}\exp\left(\jj 2\pi \frac{\rho^{\rm AP}_{m,n}\left(x,y\right)}{c}f\right),
\end{equation}
where $\rho^{\rm AP}_{m,n}\left(x,y\right)$ has been defined in~\eqref{eq:d1mn}.
% In the case we want to align the BS transmitted signal phase at frequency $f \in \Fc$ on the IRS at point $\pv=\gv(f) \in \Zc$, the ideal component of the steering matrix applied to the $\left[m,n\right]$ element of the BS array would be
% \begin{equation}
% \label{eq:IdealSteer}
% \left[\Vm\left(f\right)\right]_{mn} = \exp\left(-\jj 2\pi \frac{d_{1,mn}\left(\gv\left(f\right)\right)}{c}f\right),
% \end{equation}
% where function $\gv(f)$ maps the frequency $f$ to a specific point $\pv$ on the IRS surface as a design choice. This situation constitutes an ideal case where the weights applied to the BS antenna elements are frequency dependent and can be arbitrarily set. This is not the case in practical conditions, though several ideas have been proposed to overcome this issue, such as the possibility to set true time delays \cite{Zhao24}.

The AP UPA is configured to generate a beam directed toward the IRS. Such
configuration is described by the $M \times N$ steering matrix (or
beamformer) $\Vm(f)$. We consider here several options for the beamformer design, taking into account the beam direction and how it compensates for the BS effect.
In the most general case, the $(m,n)$-th component
of $\Vm(f)$, can take the form
\begin{equation}
\label{eq:GeneralSteer}
\left[\Vm\left(f\right)\right]_{mn} = \exp\left(-\jj 2\pi \frac{\rho^{\rm AP}_{m,n}\left(\gv\left(u(f)\right)\right)}{c}u\left(f\right)\right),
\end{equation}
where the function $u\left(\cdot\right): \RR\rightarrow \RR$ is the frequency at which the beamformer is tuned for signal frequency $f$,
and the function $\gv(f'): \Fc \rightarrow \RR^2$ specifies the point on the IRS surface which the
beam at frequency $f'$ is directed towards.
The expression in~\eqref{eq:GeneralSteer} can encompass many beamforming designs, such as the following ones.
%maps frequency $f$ into another
%frequency and function $\gv\left(u\left(f\right)\right)$ maps the
%frequency $u\left(f\right)$ to a specific point $\pv$ on the IRS
%surface as a design choice. Here we can consider three possibilities:
\begin{itemize}
\item {\em Ideal beamforming}, which perfectly eliminates the BS effect. In such a case we set $u\left(f\right)=f$ for every frequency $f \in \Fc$. Also, the beam points to a fixed (i.e., frequency-independent) position on the IRS surface. In other words $\gv\left(f\right)$ is constant.
This is mathematically equivalent to eliminating the BS, e.g., using the techniques proposed in~\cite{WB_beamforming_mMIMO,Elbir_2023}.
  
\item {\em Central beamforming}, which does not compensate for BS and is
  conceived for narrowband signals, to be employed when the BS effect does
  not entail relevant performance degradation. In such a case we set
  $u(f) = f_0$, $\forall f\in \Fc$ (we recall that $f_0$ is the central
  frequency of the signal spectrum.) Clearly,
$\gv\left(u(f)\right)=\gv(f_0)$ is also a constant function.  
  
\item {\em Hybrid beamforming}, which represents a trade-off between {\em central} and {\em ideal beamforming}.
  The idea here is to organize the band $\Fc$ into $K$ narrower disjoint
 sub-bands, and to apply a different steering matrix to each
 of them. Specifically, 
  $u(f) = f_k$ for $f\in \Fc_k$, $k=1,\ldots,K$, where
 $f_k\in \Fc_k =[f_{k,\min},f_{k,\max}] \subseteq \Fc$; the sets
 $\Fc_k$ are such that $\cup_k\Fc_k=\Fc$ and
 $\Fc_{k'}\cap \Fc_k=\emptyset$ for $k\neq k'$.
\end{itemize}

Let $s(t)$ be the signal transmitted from the AP, and $S(f)$ its
spectrum. Moreover, let
\beq
\label{eq:P_TX}
P_{\rm TX} = \int_{\mathcal{F}} \left| S\left(f\right) \right|^2 df
\eeq
be the transmitted power.
Then, the spectrum of the signal impinging on the IRS at point
$\pv_{\Gc}=\left(x,y,0\right)\Tran$ is given
by
\begin{equation}
\label{eq:RecIRS}
R\left(x,y,f\right) = S\left(f\right) W\left(x,y,f\right),
\end{equation}
where 
\ifonecol
\begin{equation}
\label{eq:wfunc}
W\left(x,y,f\right) = \trace\left\{\Um(x,y,f)\Tran \Vm\left(f\right)\right\}) = \sum_{m=0}^{M-1}\sum_{n=0}^{N-1}\frac{1}{\sqrt{4\pi}d_{1,mn}\left(x,y\right)}\exp\left(\frac{\jj 2\pi}{c}\left(d_{1,mn}\left(x,y\right) f-d_{1,mn}(\gv\left(u\left(f\right)\right)\right)u\left(f\right)\right)\right),
\end{equation}
\else
\begin{eqnarray}
\label{eq:wfunc}
& \displaystyle W\left(x,y,f\right) = \trace\left\{\Um(x,y,f)\Tran \Vm\left(f\right) \right\} & \\
& \displaystyle = \sum_{m=0}^{M-1} \sum_{n=0}^{N-1}\frac{1}{\sqrt{4\pi}\rho^{\rm AP}_{m,n}\left(x,y\right)}& \nonumber \\
& \displaystyle \cdot \exp\left(\frac{\jj 2\pi}{c}\left(\rho^{\rm AP}_{m,n}\left(x,y\right) f-\rho^{\rm AP}_{m,n}\left(\gv\left(u\left(f\right)\right)\right)u\left(f\right)\right)\right).& \nonumber
\end{eqnarray}
\fi
The spectrum of the signal scattered from IRS point $\pv_{\Gc}$ and received at the UE, can be written as
\ifonecol
\begin{equation}
\label{eq:RecUE}
Z(x,y,f)  = R(x,y,f) \frac{1}{\sqrt{4\pi}d_2\left(x,y\right)} \exp\left(\jj2\pi \frac{d_2\left(x,y\right)}{c}f\right) \zeta\left(x,y,f\right),
\end{equation}
\else
\begin{eqnarray}
\label{eq:RecUE}
&\displaystyle Z(x,y,f) = R(x,y,f) \frac{1}{\sqrt{4\pi}\rho^{\rm UE}\left(x,y\right)}& \nonumber\\ 
&\displaystyle \cdot \exp\left(\jj2\pi \frac{\rho^{\rm UE}\left(x,y\right)}{c}f\right) \zeta\left(x,y,f\right),& 
\end{eqnarray}
\fi where $\zeta\left(x,y,f\right)$ is the frequency response of the
IRS at $\pv_{\Gc}$. In general, we can write~\cite{Badheka23}
\begin{equation}
\label{eq:IRSResp}
\zeta\left(x,y,f\right) =  \widetilde{\zeta}(f) \exp\left(\jj \phi\left(x,y\right)\right),
\end{equation}
where $\widetilde{\zeta}\left(f\right)$ is the uncontrollable part
of the IRS frequency response, assumed to be independent of the coordinates $(x,y)$, since generally all IRS elements are identical and are produced by the same manufacturing process. Furthermore, $\phi(x,y)$ is the
controllable phase shift applied to the signal impinging at $\pv_{\Gc}$, which, in practice, is independent of frequency. This is a typical hypothesis in most IRS literature. 
% Within the operational bandwidth of the IRS, we can consider that both parts are independent of each other, so that we can approximate the IRS frequency response by
% \begin{equation}
% \label{eq:IRSAppResp}
% \zeta\left(x,y,f\right) =  \widetilde{\zeta}\left(f\right) \exp\left(\jj \phi\left(x,y\right)\right),
% \end{equation}
% where $\widetilde{\zeta}\left(f\right)$ accounts for the frequency response regardless of the position, and $\phi\left(x,y\right)$ accounts for the IRS determined phase shift at the given point, regardless of the signal frequency.

Taking all this into account and approximating the
IRS as a continuum of scatterers, the spectrum of the 
signal received at the UE can be obtained by aggregating the signal contributions
from every point on the IRS surface $\Zc$, i.e.
\ifonecol
\begin{eqnarray}
\label{eq:TotRecUE}
Z(f) &=& \int_\Zc Z(x,y,f) \dd x\dd y \\
& = & S(f)\int_\Zc\frac{\zeta(x,y,f)\exp\left(\jj 2\pi \frac{\rho^{\rm UE}(x,y)}{c}f\right) W(x,y,f)}{\sqrt{4\pi}\rho^{\rm UE}(x,y)}\dd x \dd y, \nonumber
\end{eqnarray}
\else
\begin{eqnarray}
\label{eq:TotRecUE}
Z(f) &=& \int_\Zc Z(x,y,f) \dd x\dd y \nonumber \\
     &=&S(f) H(f)
\end{eqnarray}
\fi
where $H(f)$ is the overall system transfer function $H(f)$ defined as
\ifonecol
\begin{equation}
\label{eq:TransFunc}
H(f) \triangleq \int_\Zc \widetilde{\zeta}(x,y,f)\frac{\exp\left(\jj \varphi(x,y,f)\right)\exp\left(\jj 2 \pi \frac{\rho^{\rm UE}(x,y)}{c}f\right) W(x,y,f)}{\sqrt{4\pi}\rho^{\rm UE}(x,y)}\dd x \dd y.
\end{equation}
\else
\begin{eqnarray}
\label{eq:TransFunc}
H(f) &=&\int_\Zc \widetilde{\zeta}(f) \exp\left(\jj \phi(x,y)\right) \nonumber \\
&&\qquad \cdot \frac{\exp\left(\jj 2\pi \frac{\rho^{\rm UE}(x,y)}{c}f\right) W(x,y,f)}{\sqrt{4\pi}\rho^{\rm UE}(x,y)}\dd x \dd y. \nonumber \\
& &
\end{eqnarray}
\fi
Finally, the total received power is given by
\beq \label{eq:P_RX}
P_{\rm RX} = \int_{\mathcal{F}} \left| Z\left(f\right) \right|^2 df\,.
\eeq

\subsection{Problem statement and roadmap}
%{\color{red} Here we need to clearly state the problem. Maximizing the
%transfer function is a bit misleading. We need to find a metric to
%maximize. Sometimes we maximize the transfer function, sometimes the %received power}  
%Now the typical problem is to find suitable phase shifts
%$\phi(x,y,f)$ in order to coherently recombine at the UE the signal
%scattered from the IRS and maximize the transfer function. To tackle
%this, in the following we consider different situations and provide
%the optimal solution and a number of approximations.
The effect of the IRS on the quality of the communication link depends on its configuration, i.e., on the setting of the phase shifts $\phi(x,y)$. In the following sections, we will consider several possible IRS configuration techniques, which we will compare on the basis of several metrics. We also assume that the AP adopts the central beamforming technique, i.e., we consider the most challenging scenario where the impinging signal is fully affected by BS. Indeed, ideal and hybrid beamforming techniques partially or completely remove BS at the source, at the price of increased AP hardware and software complexity, leaving less room for performance improvements through IRS phase shift optimization.   

In particular, as done in~\cite{Carlson22}, whose setup was quite similar to the one considered here, we adopt as a metric of interest the power received at the UE, normalized with respect to the signal bandwidth, given by
\begin{equation}
\label{eq:Z_avg}
\Pc\left(B\right) = \frac{P_{\rm RX}}{B}\, ,
\end{equation}
and we will refer to it as the average received power spectral density. Accordingly, the best IRS configuration is the one that maximizes the above metric, i.e.
\beq
\phi^{\rm opt}(x,y) = \arg \max_{\phi(x,y)} \Pc\left(B\right)\,.
\eeq

Another figure of merit that we shall consider is a sort of spectrum-aware standard deviation of the received spectrum, calculated as
\begin{eqnarray}
\label{eq:sig_Z}
&\displaystyle \sigma\left(B\right) = \sqrt{\frac{\int_{\mathcal{F}} \left| \left|Z\left(f\right)\right|^2 - \frac{P_{\rm RX}}{P_{\rm TX}} \left|S\left(f\right)\right|^2\right|^2 df}{B}}& \nonumber \\
&\displaystyle =\sqrt{\frac{\int_{\mathcal{F}} \left|S\left(f\right)\right|^4 \left| \left|H\left(f\right)\right|^2 - \frac{P_{\rm RX}}{P_{\rm TX}}\right|^2 df}{B}}.& %\nonumber
\end{eqnarray}

This definition, which takes into account the spectrum of the transmitted signal, provides a measurement of the distortion caused on the transmitted spectrum by the channel: a higher value in the standard deviation points towards a higher distortion.
Observe that if $|H(f)|^2$ is constant on $\Fc$ then by~\eqref{eq:TotRecUE}, \eqref{eq:P_TX} and \eqref{eq:P_RX} we have $|H(f)|^2-\frac{P_{\rm RX}}{P_{\rm TX}}=0$. Hence, there is no distortion since $\sigma(B)=0$.

In general the distortion has at least two causes: 
\begin{itemize} 
\item {\em geometric distortion} due to BS and to the IRS side length not being negligible compared to the AP--IRS and the IRS--UE distances. In such a case, two distinct signal paths connecting the AP and the UE through the IRS and impinging on two different IRS points have different lengths and, hence, they are subject to different free space attenuation. But since the angle of departure from the AP of such paths is frequency dependent, due to BS, the resulting received spectrum is also frequency dependent, i.e. it is distorted due to geometry. Note that this effect might be partially compensated for by properly adjusting the gain of the IRS elements, if this option is allowed by the IRS design.
\item IRS {\em configuration distortion} due to a specific choice of the controllable phase shifts, $\phi(x,y)$, which appear in the expression of $H(f)$ given in~\eqref{eq:TransFunc}. If $\sigma(B)$ is considered as a performance metric, an IRS configuration $\phi(x,y)$ is preferable to another if it makes $H(f)$ closer to a constant on $\Fc$.
\end{itemize}

Since different IRS configurations have different mean values $\Pc$, we will preferably compare them on the basis of the coefficient of variation \cite{Yadolah08}, defined as the normalized standard deviation, i.e., 
\begin{equation}
\label{eq:CV}
{\rm CV}\left(B\right) = \frac{\sigma\left(B\right)}{\Pc\left(B\right)}.
\end{equation}
An IRS configuration is preferable to another with a similar value of $\Pc$ if it has a lower coefficient of variation.

\section{Upper bound to $|H(f)|$\label{sec:upper_bound}}
In this section, we derive an upper bound to the magnitude of the transfer function, in the hypothesis that the controllable part of the IRS phase shifts can be arbitrarily set for any frequency $f$. Let $\phi_{\rm UB}(x,y,f)$ denote the frequency-dependent phase shift that maximizes the transfer function $|H(f)|$ in~\eqref{eq:TransFunc}, for all $f\in\Fc$. It can be obtained by imposing that all signals scattered by the IRS reach the UE with the same phase. This means that the argument within the integral in~\eqref{eq:TransFunc} should have a constant phase for all $f \in \Fc$, i.e., discarding the real terms, 
\ifonecol
\begin{equation}
\label{eq:argOpt}
\displaystyle \arg\left\{ \widetilde{\zeta}\left(x,y,f\right)\exp\left(\jj \phi_{\rm UB}\left(x,y,f\right)\right)\exp\left(\jj2\pi \frac{d_2\left(x,y\right)}{c} f\right) W\left(x,y,f\right)\right\} = K,
\end{equation}
\else
\begin{eqnarray}
\label{eq:argOpt}
&\displaystyle \arg\left\{ \widetilde{\zeta}(f)\exp\left(\jj \phi_{\rm UB}\left(x,y,f\right)\right)\exp\left(\jj2\pi \frac{\rho^{\rm UE}\left(x,y\right)}{c} f\right)\right. & \nonumber \\
&\displaystyle \cdot \Big. W\left(x,y,f\right)\Big\} = K,&
\end{eqnarray}
\fi
where $K$ is an arbitrary constant. From the above equation, we obtain that
\ifonecol
\begin{equation}
\label{eq:OptPhase}
\displaystyle \phi_{\rm OPT}\left(x,y,f\right) = -2\pi \frac{d_2\left(x,y\right)}{c} f - \arg\left\{\widetilde{\zeta}\left(x,y,f\right) W\left(x,y,f\right) \right\} + K
\end{equation}
\else
\begin{eqnarray}
\label{eq:OptPhase}
&\displaystyle \phi_{\rm UB}\left(x,y,f\right) = -2\pi \frac{\rho^{\rm UE}\left(x,y\right)}{c} f & \\
&\displaystyle - \arg\left\{\widetilde{\zeta}(f) W\left(x,y,f\right) \right\} + K & \nonumber
\end{eqnarray}
\fi
which corresponds to the transfer function
\begin{equation}
\label{eq:Hopt}
H_{\rm UB}(f)= \ee^{\jj K} \int_{\Zc} \frac{\left|\widetilde{\zeta}(f) W\left(x,y,f\right)\right|}{\sqrt{4\pi} \rho^{\rm UE}\left(x,y\right)} \dd x \dd y.
\end{equation}
Note that $|H_{\rm UB}(f)|$ is an upper bound for $\left|H\left(f\right)\right|$, because
\begin{eqnarray}
\label{eqUpBoundH}
\left|H(f)\right| &\hspace{-1ex}{=}&\left|\displaystyle \int_\Zc W\left(x,y,f\right) \frac{\zeta\left(x,y,f\right)\ee^{\jj2\pi f\rho^{\rm UE}(x,y)/c} }{\sqrt{4\pi}\rho^{\rm UE}\left(x,y\right)}\dd x \dd y \right| \nonumber \\
&\hspace{-1ex}\le&\int_\Zc\frac{\left| \widetilde{\zeta}(f) W\left(x,y,f\right)\right|}{\sqrt{4\pi}\rho^{\rm UE}\left(x,y\right)}\dd x \dd y \nonumber \\
&\hspace{-1ex}=&\left| H_{\rm UB}(f)\right|.
\end{eqnarray}

We point out that the solution in~\eqref{eq:OptPhase} is not feasible since in practical settings the controllable phase shifts do not depend on $f$.

\section{Baseline IRS configuration techniques}
\label{sec:baseline}

In this section, we summarize some IRS configuration techniques that have already been proposed in the literature and that will serve as a benchmark to measure the performance of our solutions.

\subsection{Narrowband techniques}

% Practically, the IRS phase is not freely tunable for every frequency $f$ within the considered bandwidth,  due to hardware constraints. Thus, setting the IRS phase shifts as in \eqref{eq:OptPhase} is not feasible. 
A narrowband IRS configuration, conceived for the case in which the signal bandwidth is not so large, consists in aligning in phase all the scattered signal components arriving at the UE for a single frequency $f_{\rm NB}$. In this case by particularizing~\eqref{eq:OptPhase} for $f=f_{\rm NB}$ we get
\ifonecol
\begin{equation}
\label{eq:IdealPhase_f0}
\displaystyle \phi\left(x,y,f\right) \triangleq \phi\left(x,y,f_{\rm NB}\right)^{\rm NB} = -2\pi \frac{d_2\left(x,y\right)}{c} f_{\rm NB} - \arg\left\{\widetilde{\zeta}\left(x,y,f_{\rm NB}\right) W\left(x,y,f_{\rm NB}\right)\right\} + K,
\end{equation}
\else
\begin{eqnarray}
\label{eq:IdealPhase_f0}
&\displaystyle \phi_{\rm NB}\left(x,y\right) \triangleq \phi_{\rm UB}\left(x,y,f_{\rm NB}\right) = -2\pi \frac{\rho^{\rm UE}\left(x,y\right)}{c} f_{\rm NB} & \\
&\displaystyle - \arg\left\{\widetilde{\zeta}(f_{\rm NB}) W\left(x,y,f_{\rm NB}\right)\right\} + K, & \nonumber
\end{eqnarray}
\fi
and the corresponding transfer function, $H_{\rm NB}(f)$, is obtained by setting $\phi\left(x,y\right) = \phi_{\rm NB}\left(x,y\right)$  in~\eqref{eq:TransFunc}
\ifonecol
\begin{eqnarray}
\label{eq:IdealNB_f0}
&\displaystyle H(f)^{\rm NB} \triangleq \left.H(f)\right|_{\phi(x,y,f)=\phi(x,y,f_0)^{\rm NB}}& \\
&\displaystyle = \exp\left(\jj K\right) \int_\Zc\frac{\widetilde{\zeta}\left(x,y,f\right) w\left(x,y,f\right)\exp\left(-\jj \arg\{\widetilde{\zeta}\left(x,y,f_{\rm NB}\right) w\left(x,y,f_0\right)\}\right)\exp\left(\jj 2\pi \frac{d_2\left(x,y\right)}{c} \left(f-f_{\rm NB}\right)\right)}{\sqrt{4\pi} d_2\left(x,y\right)}\dd x \dd y. \nonumber
\end{eqnarray}
\else
\begin{eqnarray}
\label{eq:IdealNB_f0}
&\displaystyle H_{\rm NB}(f) \triangleq \left.H(f)\right|_{\phi(x,y)=\phi_{\rm NB}(x,y)} = \ee^{\jj K} & \\
&\displaystyle \cdot \int_\Zc \widetilde{\zeta}(f) W\left(x,y,f\right) \exp\left(\jj 2\pi \frac{\rho^{\rm UE}\left(x,y\right)}{c} \left(f-f_{\rm NB}\right)\right) & \nonumber\\
&\displaystyle \cdot \frac{\exp\left(-\jj \arg\{\widetilde{\zeta}(f_{\rm NB}) W\left(x,y,f_{\rm NB}\right)\}\right)}{\sqrt{4\pi} \rho^{\rm UE}\left(x,y\right)}\dd x \dd y. \nonumber
\end{eqnarray}
\fi
This expression is valid for both far and near-field since we did not make any hypothesis on the relation between the AP--IRS and IRS-UE distances and IRS/AP size.
As for the choice of $f_{\rm NB}$, one could consider the central frequency of the signal spectrum $S(f)$, i.e., $f_{\rm NB}=f_0$. Alternatively, if $|S(f)|$ is known, a spectrum-aware candidate for $f_{\rm NB}$ is the barycenter
\beq
f_{\rm NB} = \frac{\int_{\Fc} f |S(f) |^2 \dd f }{\int_{\Fc} | S (f) |^2 \dd f }.
\eeq
 The above definition can be considered as the average frequency with respect to the transmitted spectrum.
%{\color{red} we need to explain why he barycenter is considered. Do we have simulations with values for $f_{\rm NB}$ other than $f_0$?.}

For narrowband signals around frequency $f_{\rm NB}$ and under far-field conditions, in the following denoted as NBF, the phase shifts $\phi(x,y)$ can be set according to the generalized Snell's law for the reflected wave~\cite{Yu11, Aieta12}, which can be written as
\ifonecol
%TBD
\else
\begin{eqnarray}
\label{eq:GenSnell}
&\displaystyle \phi_x(x,y) {=} {-}\frac{2\pi f}{c}\left[\sin \theta^{\rm UE} \cos \psi^{\rm UE}{+} \sin \theta^{\rm AP} \cos \psi^{\rm AP}\right]  & \nonumber \\
&\displaystyle \phi_y(x,y) {=}{-}\frac{2\pi f}{c}\left[\sin \theta^{\rm UE} \sin \psi^{\rm UE} {+} \sin \theta^{\rm AP} \sin \psi^{\rm AP}\right],& \nonumber \\ 
\end{eqnarray}
\fi
where $\phi_x(x,y)= \frac{\partial}{\partial x} \phi\left(x,y\right)$ and, analogously, $\phi_y(x,y)= \frac{\partial}{\partial y} \phi\left(x,y\right)$. The angles $\theta^{\rm AP}(x,y)$ and  $\psi^{\rm AP}(x,y)$ are the spherical coordinates of the AP array center, as observed from $(x,y,0)$ and depicted in Fig.~\ref{fig:irs_3D}. Similarly,  $\theta^{\rm UE}(x,y)$ and  $\psi^{\rm UE}(x,y)$ are the spherical coordinates of the UE.  In~\eqref{eq:GenSnell} we have omitted the angles' arguments for simplicity. Therefore, by particularizing for frequency $f_{\rm NB}$, we can obtain phase shifts $\phi_{\rm NBF}\left(x,y\right)$ satisfying~\eqref{eq:GenSnell} as
%\ifonecol
%TBD
%\else
%\begin{eqnarray}
%\label{eq:GenSnell_NBLDS}
%&\displaystyle \frac{\partial \phi_{\rm NBLDS}\left(x,y\right)}{\partial x} \approx -\frac{2\pi f_0}{c} & \\
%&\displaystyle \cdot \left[\sin \theta^{\rm UE}_0 \cos \psi^{\rm UE}_0 + \sin \theta^{\rm BS}_0 \cos \psi^{\rm BS}_0\right] & \nonumber \\
%&\displaystyle \frac{\partial \phi_{\rm NBLDS}\left(x,y\right)}{\partial y} \approx -\frac{2\pi f_0}{c} & \\
%&\displaystyle \cdot \left[\sin \theta^{\rm UE}_0 \sin \psi^{\rm UE}_0 +\sin \theta^{\rm BS}_0 \sin \psi^{\rm BS}_0\right], & \nonumber
%\end{eqnarray}
%\fi
% \begin{eqnarray}
% \frac{\partial \phi_{\rm NBF}(x,y)}{\partial x} &\hspace{-2ex}=& \hspace{-2ex}{-}\frac{2\pi f_{\rm NB}}{c} \left[\sin \theta^{\rm UE}_0 \cos \psi^{\rm UE}_0 {+} \sin \theta^{\rm AP}_0 \cos \psi^{\rm AP}_0\right]
% \nonumber \\
% \frac{\partial \phi_{\rm NBF}\left(x,y\right)}{\partial y} &\hspace{-2ex}=& \hspace{-2ex}{-}\frac{2\pi f_{\rm NB}}{c} \left[\sin \theta^{\rm UE}_0 \sin \psi^{\rm UE}_0 {+}\sin \theta^{\rm AP}_0 \sin \psi^{\rm AP}_0\right]\nonumber \\
% \label{eq:SnellNB}
% \end{eqnarray}
%   where $\theta^{\rm UE}_0 = \theta^{\rm UE}(\gv(f_{\rm NB}))$  and, analogously for $\theta^{\rm AP}_0$, $\phi^{\rm UE}_0$ and $\phi^{\rm AP}_0$. In this way, we get
\ifonecol
%TBD
\else
\beq
\begin{split}
\phi_{\rm NBF}&\left(x,y\right) = -\frac{2\pi x f_{\rm NB} }{c} 
\left[ \sin \theta^{\rm UE}_0 \cos \psi^{\rm UE}_0 + \sin \theta^{\rm AP}_0 \cos \psi^{\rm AP}_0\right] \\
&\displaystyle - \frac{2\pi y f_{\rm NB}}{c} \left[ \sin \theta^{\rm UE}_0 \sin \psi^{\rm UE}_0 
 + \sin \theta^{\rm AP}_0 \sin \psi^{\rm AP}_0\right] + K_1,
\end{split}
\eeq
\fi
where $K_1$ is a constant, $\theta^{\rm UE}_0 = \theta^{\rm UE}(\gv(f_{\rm NB}))$  and, analogously for $\theta^{\rm AP}_0$, $\phi^{\rm UE}_0$ and $\phi^{\rm AP}_0$.

% \footnote{Here we have assumed for simplicity that the phase of the uncontrollable IRS response $\arg \left( \widetilde{\zeta}(x,y,f_{\rm NB})\right)$ is constant. If not, the IRS phase shift must also compensate for it and becomes
% \[
% \begin{split}
% \phi_{\rm NBL}&\left(x,y\right) = -\frac{2\pi f_{\rm NB} x}{c} 
% \left[ \sin \theta^{\rm UE}_0 \cos \psi^{\rm UE}_0 + \sin \theta^{\rm BS}_0 \cos \psi^{\rm BS}_0\right] \\
% &\displaystyle - \frac{2\pi f_{\rm NB} y}{c} \left[ \sin \theta^{\rm UE}_0 \sin \psi^{\rm UE}_0 
%  + \sin \theta^{\rm BS}_0 \sin \psi^{\rm BS}_0\right] \\
%  &- \arg \left( \widetilde{\zeta}(x,y,f_{\rm NB})\right) + K,
% \end{split}
% \]

% }

\subsection{Eigenvalue decomposition (ED) technique}
From \eqref{eq:RecUE} and \eqref{eq:P_RX}, the  received power can be written as
\begin{equation}
P_{\mathrm RX} = \iiint_{\Fc,\Zc,\Zc} \hspace{-3ex}Z\left(x',y',f\right)^* Z\left(x,y,f\right) \dd x\dd y   \dd x'\dd y' \dd f\,.
\label{eq:P_RX_bis}
\end{equation}

By writing $Z\left(x,y,f\right) = \widetilde{Z}\left(x,y,f\right) \exp\left(\jj \phi(x,y)\right)$, where $\widetilde{Z}\left(x,y,f\right)$ includes all factors of $Z\left(x,y,f\right)$ that multiply the controllable part of the IRS phase response, we obtain
\begin{eqnarray}
P_{\mathrm RX} {=}\iint_{\Zc,\Zc}\ee^{-\jj \phi(x',y')}\Tc\left(x',y',x,y \right) \ee^{\jj \phi(x,y)}\dd x\dd y   \dd x'\dd y' \nonumber \\
\label{eq:P_RX_1}
\end{eqnarray}
where we have defined the continuous function
\beq
\Tc\left(x',y',x,y \right) = \int_{\Fc} \widetilde{Z}\left(x',y',f\right)^* \widetilde{Z}\left(x,y,f\right) \dd f
\eeq

A heuristic solution for the maximization of \eqref{eq:P_RX_1} can be obtained as follows.  We first solve the following looser problem:
\beq
\begin{split}
\gamma^*(x,y) = &\max_{\gamma(x,y),  \| \gamma\|_{\infty} = 1} \\
\int_\Zc \int_\Zc &\gamma(x',y')^*\Tc\left(x',y',x,y \right) \gamma(x,y)\dd x\dd y   \dd x'\dd y'
\end{split}
\eeq
whose solution consists in equating $\gamma^*(x,y)$ to the unit-norm eigenvector corresponding to the maximum eigenvalue of the operator\footnote{Since $\Lc$ is a self-adjoint linear bounded operator,  the supremum of all eigenvalues is bounded by the operator norm, which is finite.  If there is not a maximum eigenvalue,  we can choose an eigenvalue infinitesimally close to the supremum. In practice, we will allways work with discretized IRSs, so that the operator becomes a matrix, for which the maximum eigenvalue is always well defined.} 
\beq
\left(\Lc \gamma \right)(x',y') = \int_\Zc \Tc\left(x',y',x,y \right) \gamma(x,y) \dd x \dd y
\eeq
Then,  we set $\phi_{\rm ED}(x,y) = \arg \left( \gamma^*(x,y) \right)$.

This eigenvalue-based technique was already introduced in~\cite{Tarable2025, NoiICC}, where it is formulated in the more common discrete-element setting, and is considered here as a benchmark.  It is suboptimal since, in general,  $\gamma^*(x,y)$ does not have constant modulus (or, equivalently,  $\exp\left(\jj \phi_{\rm ED}(x,y)\right)$ is not equal to  $\gamma^*(x,y)$ up to a multiplicative constant). However, it typically has quite good performance.

\section{Local optimization techniques}
\label{sec:local}

When the central beamformer is adopted at the AP, different points of the IRS generally see a different spectrum of the impinging signal because of BS at the AP. 
In this section, we leverage this consideration by proposing IRS configuration techniques that, at each point, tune their response on the locally most significant component of the signal spectrum. 
This local IRS optimization approach will be shown in Section \ref{sec:results} to have better performance than narrowband techniques, and to be competitive with the more complex ED technique.

We consider the generalized Snell's law~\eqref{eq:GenSnell} applied to the signal reflected at generic point $\left(x,y,0\right)$ of the IRS and at frequency $f$.
The equations of the generalized Snell's law state that, at that IRS point, the signal at frequency $f$ is reflected towards the UE. In general, we might be interested in applying~\eqref{eq:GenSnell} to different frequencies at different IRS points. For this reason, we now replace $f$ in~\eqref{eq:GenSnell} with the function $f(x,y)$, defined over $(x,y,0) \in \Zc$ and whose image is $\Fc$.
The rationale is that the IRS is not uniformly illuminated over the whole spectrum of the transmitted signal, and there are points of the IRS which are reached by certain spectrum components with a larger power than others.  Thus, roughly speaking,  $f(x,y)$ should be a frequency component that reaches point $(x,y,0)$ with relevant power
\footnote{For narrowband signals whose central frequency is $f_{\rm NB}$, it all boils down to $f(x,y)=f_{\rm NB}$, $\forall (x,y,0) \in \Zc$.}.

In general, however, the system of equations in~\eqref{eq:GenSnell} is valid only if it satisfies the Schwartz conditions 
\[ \frac{\partial }{\partial x}\phi_y(x,y) = \frac{\partial }{\partial y} \phi_x(x,y)\,.\]

Since this is not the case for general $f(x,y)$,  we resort to a least-square (LS) approach to derive the phase shifts applied at the IRS elements. In particular,  let 
\[
\nabla \phi^{\mathrm{ID}}(x,y) = \left(\phi_x^{\mathrm{ID}}(x,y), \phi_y^{\mathrm{ID}}(x,y) \right)
\] 
be the ideal gradient obtained from the generalized Snell's law in~\eqref{eq:GenSnell} for the chosen frequency function $f(x,y)$. Then, to obtain the actual phase shifts that best approximate the above gradient, we solve the following quadratic-programming problem:
\beq \label{eq_qp}
\phi^*(x,y) = \arg \min_{\phi(x,y)} \|\nabla \phi - \nabla \phi^{\mathrm{ID}} \|_2^2\,.
\eeq

For what concerns the complexity of the local optimization techniques, it boils down to make $O\left(\frac{L_x L_y}{\Delta^2} \right)$ optimizations (one per each IRS element) of the value of $f(x,y)$. Additionally, the complexity to solve the  problem in \eqref{eq_qp} is at most quadratic in the number of IRS elements.
In the following subsections, we propose two different options for the choice of $f(x,y)$ under this framework.

\subsection{Spectrum-aware local optimization (SLO)} 

The signal $R\left(x,y,f\right)$ impinging on the IRS is characterized in \eqref{eq:RecIRS}. Consider a sliding window of width $w$ in the frequency domain and define
\ifonecol
%TBD
\else
\begin{eqnarray}
&\displaystyle Q_w\left(x,y,\nu\right) = \int_{\nu-w/2}^{\nu +w/2} \left|R\left(x,y,f\right)\right|^2 \dd f& \nonumber \\
&\displaystyle = \int_{\nu-w/2}^{\nu +w/2} \left|S\left(f\right)\right|^2 \left|W\left(x,y,f\right)\right|^2 \dd f.
\end{eqnarray}
\fi
Now, for every point $\left(x,y,0\right)$ on the IRS, we numerically compute
\ifonecol
%TBD
\else
\begin{equation}
f_{\rm SLO}\left(x,y\right) = \underset{\nu}{\arg \max} \,\, Q_w\left(x,y,\nu\right),
\end{equation}
\fi
In spectrum-aware local optimization, we replace $f$ in~\eqref{eq:GenSnell} with $f_{\rm SLO}\left(x,y\right)$.

\subsection{Approximate local optimization (ALO)}

Another option for the choice of $f(x,y)$ can be the one that maximizes the channel power transfer function, i.e.
\ifonecol
%TBD
\else
\begin{equation}
\label{eq:wfunc_fmax}
f(x,y) = f^*\left(x,y\right) = \underset{f}{\arg \max} \left|W(x,y,f)\right|^2.
\end{equation}
\fi
Under the assumption that the central beamforming technique is selected at the AP, with the beam at frequency $f_0$ directed toward the IRS center, i.e., $u\left(f\right)=f_0$ and $\gv\left(u\left(f\right)\right)=\left(0,0\right)$ in \eqref{eq:wfunc}, let us define
\ifonecol
%TBD
\else
\begin{equation}
 \displaystyle \eta_{mn}=\frac{1}{\sqrt{4\pi}\rho^{\rm AP}_{m,n}\left(x,y\right)},
\end{equation}
\fi
and
\ifonecol
%TBD
\else
\begin{equation}
 \displaystyle \xi_{mn}\left(f\right)=\frac{2\pi}{c}\left(\rho^{\rm AP}_{m,n}\left(x,y\right) f-\rho^{\rm AP}_{m,n}\left(0,0\right) f_0\right),
\end{equation}
\fi
where we have omitted the dependence on $x$ and $y$ to simplify the notation. According to this, we can write
\ifonecol
%TBD
\else
\begin{equation}
\label{eq:wfunc_sq}
 \displaystyle \left|W\left(x,y,f\right)\right|^2=\!\!\!\sum_{m,n,m',n'}\!\!\!\! \eta_{mn} \eta_{m'n'} \cos\left(\xi_{mn}\left(f\right)-\xi_{m'n'}\left(f\right)\right),
\end{equation}
\fi
where the summation extends for $m=0,\cdots,M_{{\rm BS}-1}$, $n=0,\cdots,N_{{\rm BS}-1}$, $m'=0,\cdots,M_{{\rm BS}-1}$, $n'=0,\cdots,N_{{\rm BS}-1}$. If the phases $\xi_{mn}\left(f\right)$ could be freely chosen, the maximum would be reached when $\xi_{mn}\left(f\right)=\xi_{m'n'}\left(f\right)$, $\forall m,n,m',n'$. Although in general this is not possible, it is reasonable to suppose that, for $f=f^*\left(x,y\right)$, the argument of the cosine in \eqref{eq:wfunc_sq} is small, so that we can approximate it by its second-order Taylor series, yielding
\ifonecol
%TBD
\else
\begin{eqnarray}
&&\left|W\left(x,y,f\right)\right|^2 \approx \left| \sum_{m,n} \eta_{mn} \right|^2  \nonumber \\ 
&&\qquad -\frac{1}{2} \sum_{m,n,m',n'} \eta_{mn} \eta_{m'n'}\left(\xi_{mn}\left(f\right)-\xi_{m'n'}\left(f\right)\right)^2\,. \nonumber
\end{eqnarray}
\fi
In this way, problem \eqref{eq:wfunc_fmax} can be approximated by solving 
\ifonecol
%TBD
\else
\begin{equation}
\label{eq:wfunc_fmin}
 f_{\rm ALO}\left(x,y\right) = \underset{f}{\arg \min}\!\!\!\! \sum_{m,n,m',n'}\!\!\!\! \eta_{mn} \eta_{m'n'}\!\left(\xi_{mn}\left(f\right)-\xi_{m'n'}\!\left(f\right)\right)^2
\end{equation}
\fi
whose analytical solution is given by
\ifonecol
%TBD
\else
%\begin{eqnarray}
%\label{eq:f_ast}
 %&\displaystyle \widetilde{f}\left(x,y\right) = f_0& \\ 
 %&\displaystyle \cdot \sum_{m,n,m'n'} \left[ \gamma_{mn} \gamma_{m'n'} %\left( \rho^{\rm AP}_{m,n}\left(x,y\right) - \rho^{\rm AP}_{m',n'}\left(x,y\right)\right) \right. & \nonumber \\
 %&\displaystyle \cdot\left.\left( \rho^{\rm AP}_{m,n}\left(0,0\right) - \rho^{\rm AP}_{m',n'}\left(0,0\right)\right)\right] & \nonumber \\
 %&\displaystyle / \sum_{m,n,m'n'} \gamma_{mn} \gamma_{m'n'} \left( \rho^{\rm AP}_{m,n}\left(x,y\right) - \rho^{\rm AP}_{m',n'}\left(x,y\right)\right)^2.& \nonumber
%\end{eqnarray}
\beq
\begin{split}
f_{\rm ALO}&\left(x,y\right) = f_0 \,\,\times \\
&\frac{\displaystyle \sum_{m,n,m'n'}  \eta_{mn} \eta_{m'n'} \Delta\rho^{\rm AP}_{m,n,m',n'}\left(x,y\right)\Delta\rho^{\rm AP}_{m,n,m',n'}\left(0,0\right)}{\displaystyle\sum_{m,n,m'n'}  \eta_{mn} \eta_{m'n'} \left( \Delta\rho^{\rm AP}_{m,n,m',n'}\left(x,y\right)\right)^2}    
\end{split}
\eeq
\fi
where we have defined
\beq
\Delta\rho^{\rm AP}_{m,n,m',n'}\left(x,y\right) = \rho^{\rm AP}_{m,n}\left(x,y\right) - \rho^{\rm AP}_{m',n'}\left(x,y\right)\,.
\eeq
%Finally, we can write
%\ifonecol
%TBD
%\else
%\begin{eqnarray}
%\label{eq:GenSnell_appx}
%&\displaystyle \frac{\partial %\phi\left(x,y,\widetilde{f}\left(x,y\right)\right)^{\rm APP}}{\partial x} = -\frac{2\pi \widetilde{f}\left(x,y\right)}{c} & \\
%&\displaystyle \cdot\left[\sin \theta^{\rm UE} \cos \psi^{\rm UE} + \sin \theta^{\rm BS} \cos \psi^{\rm BS}\right]  & \nonumber \\
%\label{eq:GenSnell_appy}
%&\displaystyle \frac{\partial %\phi\left(x,y,\widetilde{f}\left(x,y\right)\right)^{\rm APP}}{\partial y} = -\frac{2\pi \widetilde{f}\left(x,y\right)}{c} & \\
%&\displaystyle \cdot\left[\sin \theta^{\rm UE} \sin \psi^{\rm UE} + \sin \theta^{\rm BS} \sin \psi^{\rm BS}\right],& \nonumber
%\end{eqnarray}
%\fi
%where again we have omitted the dependence of the angles with $x$ and $y$ to ease notation.

In summary, the ALO method consists in substituting $f = f_{\rm ALO}\left(x,y\right)$ into \eqref{eq:GenSnell}. To make ALO spectrum-aware, we can define $f$ in \eqref{eq:GenSnell} as follows:
\beq \label{eq:f_ALO}
f = \left\{
\begin{array}{ll}
  f_{\rm ALO}\left(x,y\right),   & \mbox{if }  f_{\rm ALO}\left(x,y\right) \in \Fc,\\
  \arg \min_{f' \in \Fc} | f'  - f_{\rm ALO}\left(x,y\right)|,   & \mbox{otherwise}.
\end{array}
\right.
\eeq

\section{Simulation results and discussion}
\label{sec:results}

% {\color{red}[Beware: as spoken in the last meeting, it would be interesting providing results for signals with non-flat spectra. This means that the metrics that make sense should be calculated over $\left|H\left(f\right)\right|^2$ instead of over $\left|Z\left(f\right)\right|^2$ (specially the standard deviation). Spectra plots still could be meaningful with $\left|Z\left(f\right)\right|^2$.

% Other thing to do with the software: create a version working with equal $\Delta f$ for every value of $B$ (current one works with equal number of frequency bins for every value of $B$).]}

We now report the performance of the proposed IRS configuration techniques, for various system settings, taking into account the IRS size, the overall signal bandwidth, the spectrum shape, and the possibility to have the bandwidth split into different bands separated by small gaps. The analysis will highlight advantages and disadvantages of each proposed technique, while illustrating significant trade-offs.

In our tests, the main system setup is described in Table~\ref{TableParam}. In order to calculate metrics~\eqref{eq:Z_avg} and~\eqref{eq:CV} as functions of bandwidth $B$, we will consider values of $B$ ranging from $0.1 f_0$ to $0.4 f_0$ (relative bandwidth from $10\%$ to $40\%$), around central frequency $f_0$. In every case, the transmitted signal power is conventionally normalized to unity, and the integrals on the frequency variable $f$ are performed numerically, using 100 discretization steps.

\begin{table}[!ht]
\begin{center}
\begin{tabular}{|c|c|}
\hline
Parameter & Value \\
\hline
$f_0$ (GHz) & $100$\\
\hline
$\av_{\Gc}$ (m) & $\left(0, -2, 1\right)\Tran$\\
\hline
$\uv_{\Gc}$ (m) & $\left(0, 1, 2\right)\Tran$\\
\hline
$M \times N$ & $64 \times 4$\\
\hline
$\alpha$ & $0^o$ \\
$\beta$ & $0^o$ \\
$\gamma$ & $60^o$ \\
\hline
$L_y$ (m) & $1.0$\\ 
\hline
\end{tabular}
\end{center}
\caption{Default parameter values}
\label{TableParam}
\end{table}

\begin{figure}[t]
\centerline{\includegraphics[width=\columnwidth]{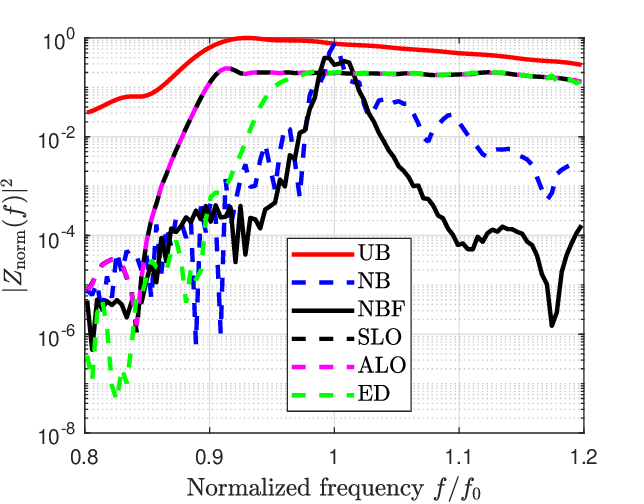}} 
\caption{Normalized received power spectrum plotted versus the normalized frequency $f/f_0$, for a single-band flat transmitted spectrum, $B = 0.4 f_0$, and  $L_x=0.2$~m.}
\label{fig:Z2_1}
\end{figure}

\begin{figure}[t]
\centerline{\includegraphics[width=\columnwidth]{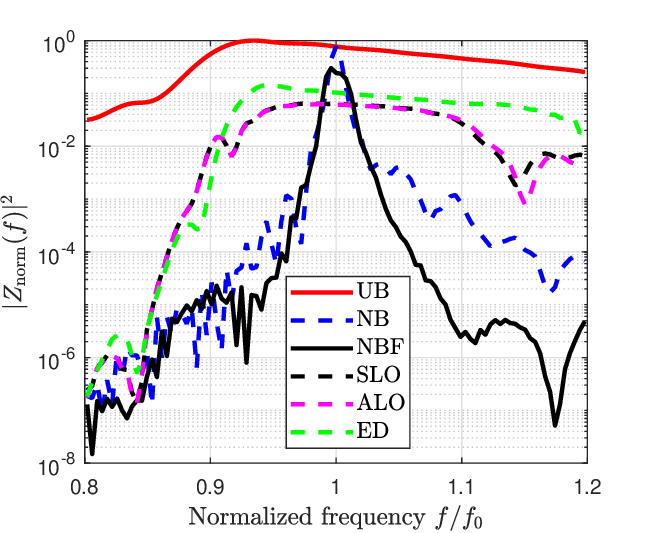}} 
\caption{Normalized received power spectrum plotted versus the normalized frequency $f/f_0$, for a single-band flat transmitted spectrum, $B = 0.4 f_0$, and  $L_x=1.0$~m.}
\label{fig:Z2_2}
\end{figure}

% {\color{red} Why $L_y/2$ instead of simply $L_y$?}
Figs.~\ref{fig:Z2_1} and~\ref{fig:Z2_2} compare the behavior of the proposed techniques for $B=0.4 f_0$ when $L_x=0.2$ m and $L_x=1.0$ m, respectively. In both cases, the spectrum of the transmitted signal, $S\left(f\right)$, is flat, i.e., constant across the entire signal bandwidth.

To ensure a fair comparison among the proposed techniques, we normalize the received signal power spectrum, $|Z\left(f\right)|^2$, with respect to the maximum of the UB response. More precisely, if $Z_{\rm UB}(f)$ denotes the received spectrum when the controllable part of the IRS response is set to $\phi_{\rm UB}$  (see Section~\ref{sec:upper_bound}), we define
\begin{equation}
\left|Z_{\rm{norm}}\left(f\right)\right|^2 = \left|Z\left(f\right)\right|^2 / \max_{f \in \Fc} \left|Z_{\rm UB}\left(f\right)\right|^2
\end{equation}
as an appropriate performance figure.

We can observe that, in general, the SLO, ALO\footnote{The ALO method is based on~\eqref{eq:f_ALO}.} and ED techniques perform better than the narrowband ones (NB, NBF), since the latter are only adapted to the barycentric frequency $f_{\rm NB} = f_0$. Indeed, SLO and ALO provide an almost flat frequency response in the range $[0.9f_0 - 1.2f_0]$, which is a highly desirable feature, as it ensures almost no spectrum distortion in that frequency range. ED performs similarly to SLO and ALO, although the flat response is reduced to $[0.97f_0-1.2f_0]$. Instead, the narrowband techniques provide a very limited pass band, about $0.03 f_0$.

For a larger IRS (as in Fig.~\ref{fig:Z2_2}), SLO and ALO are instead  outperformed by ED. This points to the fact that the ED technique is able to profit from the availability of more IRS elements, whereas SLO and ALO fail to adequately handle the growing self-interference determined by them. This is in accordance with the local optimization nature of the SLO and ALO techniques, in contrast to the global optimization performed by ED. 

In both experiments all techniques (as well as the upper bound) provide very poor performance in the range $[0.8f_0- 0.9f_0]$, but this is due to geometric reasons. Indeed, due to BS at the AP, frequencies in that range are mainly focused outside the IRS surface, so that they are not reflected toward the UE.  

\begin{figure}[t]
\centerline{\includegraphics[width=0.95\columnwidth]{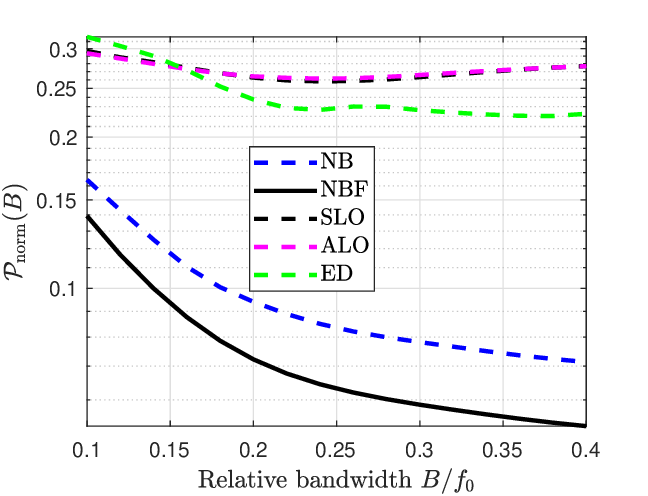}} 
\caption{Normalized average received power spectral density plotted versus the relative bandwidth $B/f_0$, for $L_x=0.2$ and a  signal characterized by a single-band flat spectrum.}
\label{fig:Zavg_1}
\end{figure}

\begin{figure}[t]
\centerline{\includegraphics[width=0.95\columnwidth]{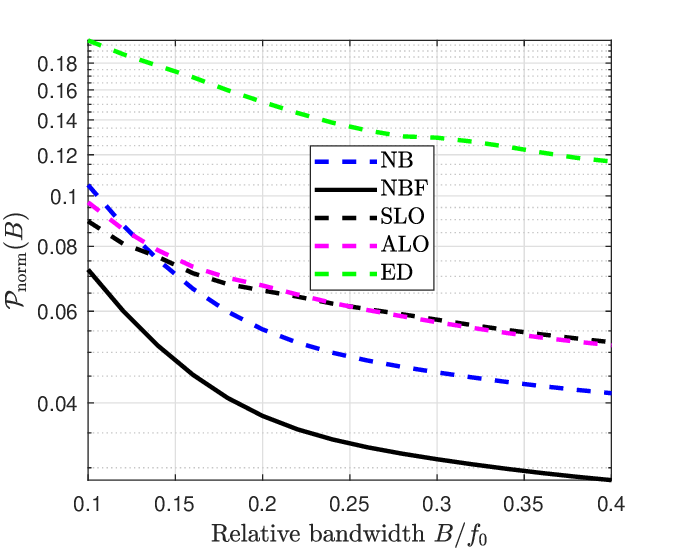}} 
\caption{Normalized average received power spectral density plotted versus the relative bandwidth $B/f_0$, for $L_x=1.0$ and a  signal characterized by a single-band flat spectrum.}
\label{fig:Zavg_2}
\end{figure}

In Figs.~\ref{fig:Zavg_1} and~\ref{fig:Zavg_2} we depict the normalized value of the average received power spectral density, $\Pc\left(B\right)$, as a function of the signal band $B$, for the same system setup used in Figs.~\ref{fig:Z2_1} and~\ref{fig:Z2_2}, respectively. For any given techniques, the normalized value of $\Pc\left(B\right)$  is defined as
\begin{equation}
\label{eq:norm_Z_avg}
\Pc_{\rm{norm}}\left(B\right)=\frac{\Pc\left(B\right)}{\Pc_{\rm{UB}}\left(B\right)},
\end{equation}
where $\Pc_{\rm{UB}}\left(B\right)$ is obtained from~\eqref{eq:Z_avg} 
when the controllable part of the IRS response is set to $\phi_{\rm UB}$.  Given this definition, in Figs~\ref{fig:Zavg_1} and~\ref{fig:Zavg_2}, the curve for $\Pc_{\rm{norm}}$ obtained for $\phi(x,y) = \phi_{\rm UB}(x,y)$ is not depicted since it is identically equal to 1. We can observe that ED provides superior performance for large IRS, whereas SLO and ALO slightly outperform ED for small IRS (i.e., when $L_x=0.2$ m) but only for bandwidths above $0.15 f_0$. As expected, the narrowband techniques provide very poor performance, with NBF being the least efficient one. For NBF this could be related to the fact that the far-field hypothesis is not met.
A general decrease of $\Pc_{\rm{norm}}\left(B\right)$ with $B$ can be observed, especially in Fig.~\ref{fig:Zavg_2}, due to the fact that the upper bound provides a flatter response for increasing bandwidth than the practical techniques.

\begin{figure}[t]
\centerline{\includegraphics[width=0.95\columnwidth]{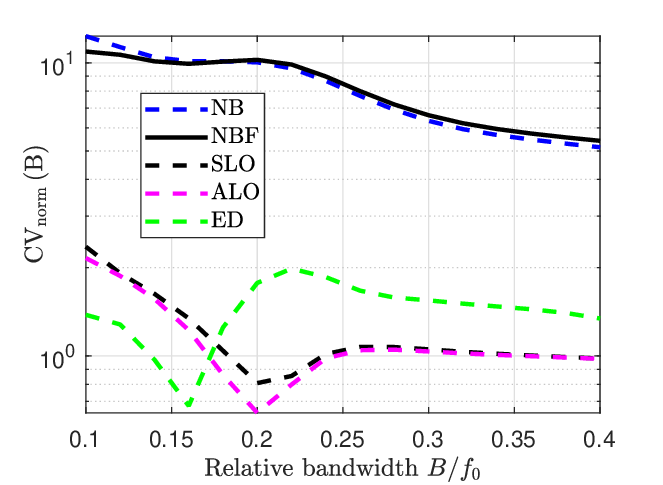}} 
\caption{Normalized coefficient of variation plotted versus the relative bandwidth $B/f_0$, for $L_x=0.2$ m and a signal  characterized by a single-band flat spectrum.}
\label{fig:SigZ_1}
\end{figure}

\begin{figure}[t]
\centerline{\includegraphics[width=0.95\columnwidth]{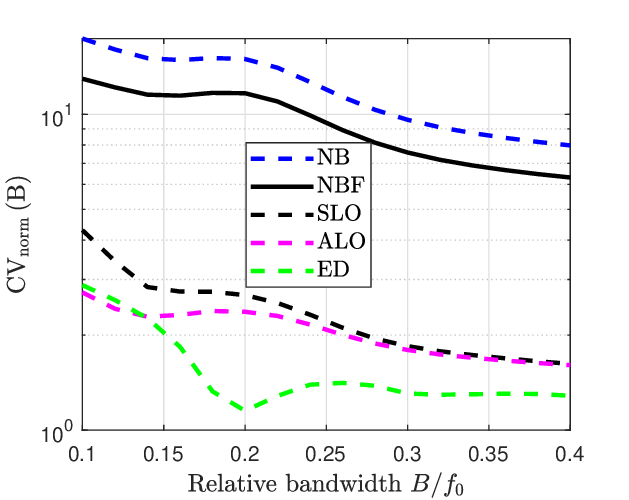}} 
\caption{Normalized coefficient of variation plotted versus the relative bandwidth $B/f_0$, for $L_x=1.0$ m and a signal  characterized by a single-band flat spectrum.}
\label{fig:SigZ_2}
\end{figure}

For the same scenarios previously considered, in Figs.~\ref{fig:SigZ_1} and~\ref{fig:SigZ_2} we depict the normalized value of the coefficient of variation given in~\eqref{eq:CV} defined as
\begin{equation}
\label{eq:norm_CV}
\rm{CV}_{\rm{norm}}\left(B\right)=\frac{\rm{CV}\left(B\right)}{\rm{CV}_{\rm{UB}}\left(B\right)},
\end{equation}
where $\rm{CV}_{\rm{UB}}\left(B\right)$ is the value of~\eqref{eq:CV} obtained when $\phi(x,y)=\phi_{\rm UB}(x,y)$. Low values of the normalized coefficient of variation arise when the system frequency response $|H(f)|$ is essentially flat over the occupied signal bandwidth. In this case also, the performance of the SLO and ALO techniques is better than that achieved by ED in the case of small IRS, and for bandwidth $B>0.18 f_0$, whereas ED outperforms all the others for large IRS. Instead, NB and NBF, which provide higher $CV_{\rm norm}$ values, lead to severe distortions of the received spectrum.

It should be noted that, in the case of Fig.~\ref{fig:SigZ_1},  SLO, ALO and ED can even outperform UB (not shown) since they achieve values smaller than unity, especially for smaller relative bandwidths. This is not a contradiction, since UB maximizes the overall power transfer, rather than the shape of the frequency response. 

%The following results we will focus on an IRS with $L_x=0.2$ m.

\begin{figure}[t]
\centerline{\includegraphics[width=0.95\columnwidth]{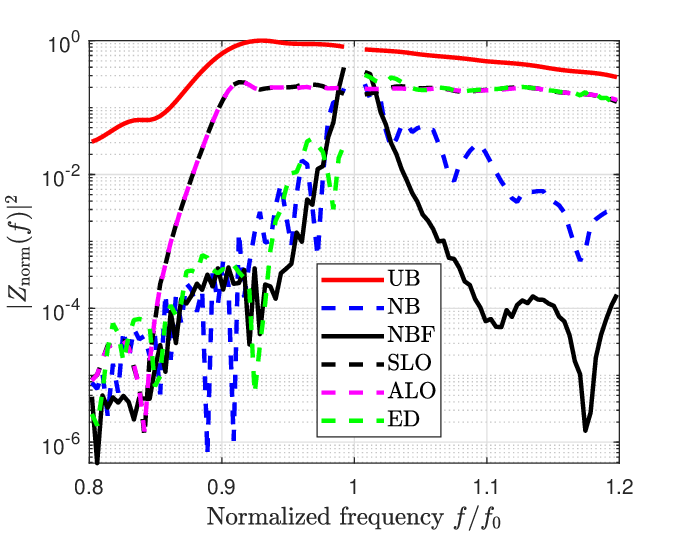}} 
\caption{Normalized received power spectrum plotted versus the normalized frequency $f/f_0$, for $L_x=0.2$ m and $S(f)$ composed of two equal-width flat bands separated by a small gap.}
\label{fig:Z2_2b}
\end{figure}

\begin{figure}[t]
\centerline{\includegraphics[width=0.95\columnwidth]{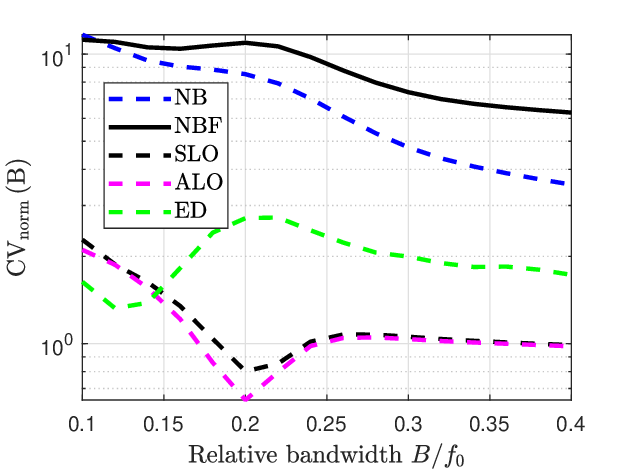}} 
\caption{Normalized coefficient of variation plotted versus the relative bandwidth $B/f_0$, for $L_x=0.2$ m and $S(f)$ composed of two equal-width flat bands separated by a small gap.}
\label{fig:sigZ_2b}
\end{figure}

In Figs.~\ref{fig:Z2_2b} and~\ref{fig:sigZ_2b}, we show the performance of the proposed techniques when the signal spectrum $S(f)$ is flat, but is composed by $2$ sub-bands of equal width, separated by a small gap, located around $f=f_0$. In Figure~\ref{fig:Z2_2b} the curves are not shown in the frequency range corresponding to the gap, since they are not defined there. The signal spectrum is depicted for $B/f_0=0.4$. Fig.~\ref{fig:sigZ_2b} shows the normalized coefficient of variation across the test bandwidth range. The normalized average received power spectral density follows a trend similar to what has already been seen in Fig.~\ref{fig:Zavg_1}.

\begin{figure}[t]
\centerline{\includegraphics[width=0.95\columnwidth]{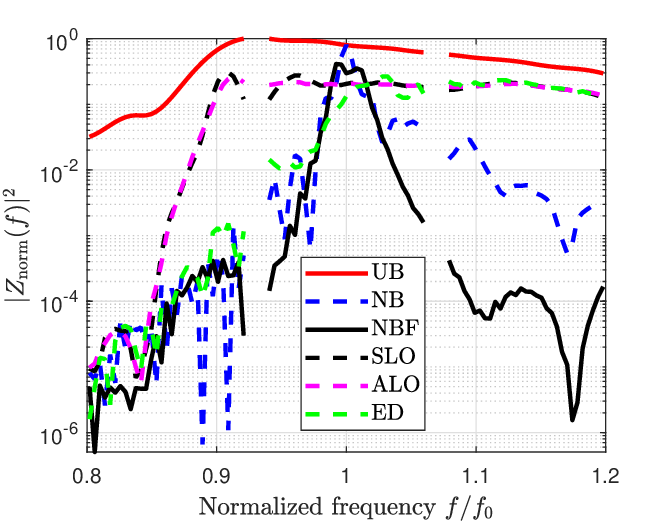}} 
\caption{Normalized received power spectrum plotted versus the normalized frequency $f/f_0$, for $L_x=0.2$ m and $S(f)$ composed of three flat sub-bands of same width, separated by small gaps.}
\label{fig:Z2_3b}
\end{figure}

\begin{figure}[t]
\centerline{\includegraphics[width=0.95\columnwidth]{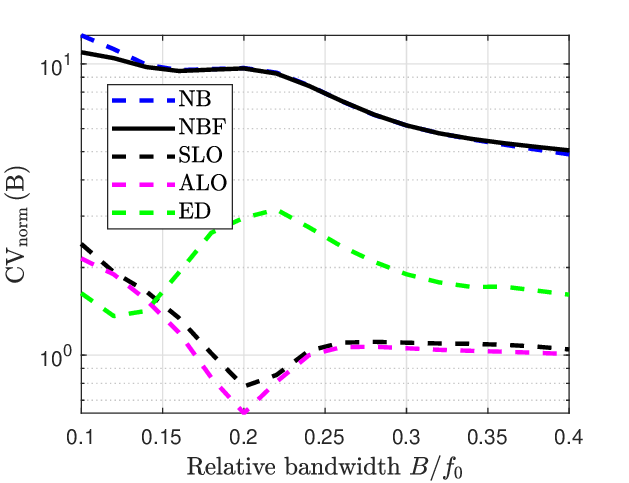}} 
\caption{Normalized coefficient of variation plotted versus the relative bandwidth $B/f_0$, for $L_x=0.2$ m and $S(f)$ composed of three sub-bands of same width, separated by small gaps.}
\label{fig:sigZ_3b}
\end{figure}

Similarly, in Figs.~\ref{fig:Z2_3b} and~\ref{fig:sigZ_3b}, we show the performance of the techniques when the signal spectrum is flat and composed of three sub-bands having the same width. In Figure~\ref{fig:Z2_3b} the curves are not shown in the frequency ranges corresponding to the gaps, since they are not defined there. Noticeably, if we focus on Fig.~\ref{fig:Z2_3b}, we observe that the ED technique fails to provide acceptable power transfer, for signals in the lower and central sub-bands, and only allows the system to operate in the upper sub-band. Instead SLO and ALO perform much better, by allowing significant power transfer in  both the central and upper sub-bands. We recall that, as discussed earlier, the frequency range $[0.8f_0, 0.9f_0]$, corresponding to most of the lower sub-band, is already penalized by the system geometry.  

As for the normalized coefficient of variation (see Fig.~\ref{fig:sigZ_3b}), observe that the performance of all techniques is almost  identical to that achieved in the case of two sub-bands (Fig.~\ref{fig:sigZ_2b}). However, above a given bandwidth (approximately $0.14 f_0$), both SLO and ALO outperform ED.

\begin{figure}[t]
\centerline{\includegraphics[width=0.95\columnwidth]{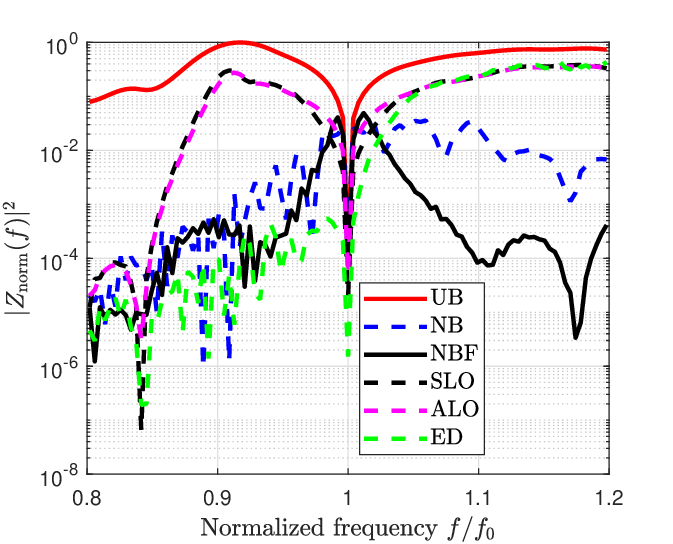}} 
\caption{Normalized received power spectrum plotted versus the normalized frequency $f/f_0$, for $L_x=0.2$ m and $|S(f)|^2$ given by~\eqref{eq:trdw}.}
\label{fig:Z2_1b_trdw}
\end{figure}

\begin{figure}[t]
\centerline{\includegraphics[width=0.95\columnwidth]{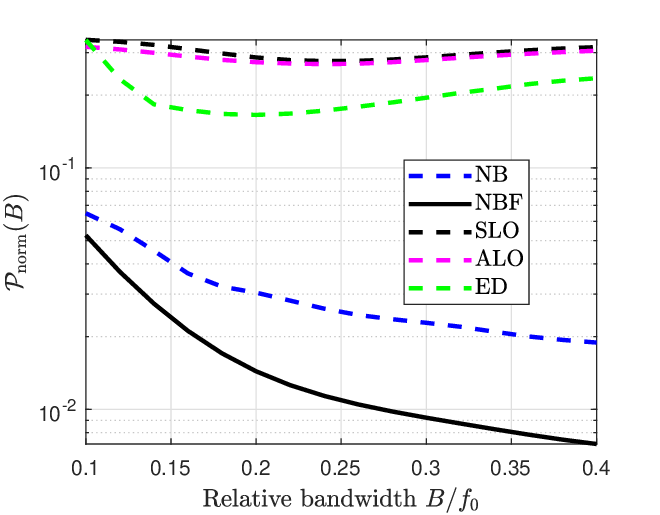}} 
\caption{Normalized average received power spectral density plotted versus $B/f_0$, for $L_x=0.2$ m and $|S(f)|^2$ given in~\eqref{eq:trdw}.}
\label{fig:Zavg_1b_trdw}
\end{figure}

\begin{figure}[t]
\centerline{\includegraphics[width=0.95\columnwidth]{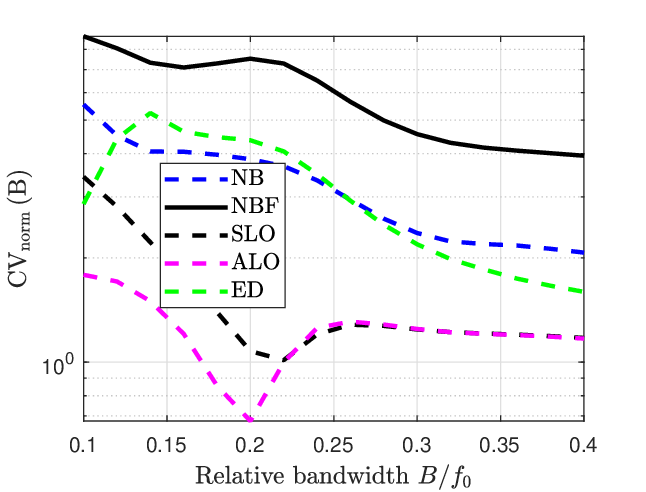}} 
\caption{Normalized coefficient of variation plotted versus $b/f_0$, for $L_x=0.2$ m and $|S(f)|^2$ given by~\eqref{eq:trdw}.}
\label{fig:sigZ_1b_trdw}
\end{figure}

Next, we analyze the case where the signal spectrum is not flat. In particular, we consider a power spectrum with shape characterized by 
\begin{equation}
\label{eq:trdw}
\left|S\left(f\right)\right|^2 = 4 \frac{|f-f_0|}{B^2}
\end{equation}
for $f \in \Ic_B$. Fig.~\ref{fig:Z2_1b_trdw} shows that ED provides poor performance in the lower half of the signal bandwidth, even lower than that provided by the narrowband techniques NB and NBF. In contrast, SLO and ALO provide excellent performance, except for $f<0.9f_0$ where  for geometric reasons, the signal focuses outside the IRS surface and cannot be reflected towards the UE. As for the normalized average received power spectral density (see Fig.~\ref{fig:Zavg_1b_trdw}), SLO and ALO outperform ED over the entire bandwidth range. Also the normalized coefficient of variation (see Fig.~\ref{fig:sigZ_1b_trdw}) shows a clear advantage of SLO and ALO (with ALO being preferred) with respect to ED over the entire bandwidth.

Overall, the provided tests show that the SLO and ALO techniques perform better than ED for small IRS, especially when the signal is composed of  more than one band or it is not flat over the frequency range. Also, we have not found significant differences between the performance offered by SLO and ALO, so that the selection of a specific method in a given context could reasonably be guided by complexity considerations or resource requirements.

\section{Conclusions}
\label{sec:conclusions}
In this work, we have reviewed and proposed spectrum-aware IRS configuration techniques that allow the communication between an AP UPA and a single antenna UE in an environment characterized by blockage of the AP--UE LoS paths. 
Such techniques take into account for the shape of the transmitted signal spectrum in the calculation of the controllable phase shift $\phi(x,y)$ to be applied to the IRS elements.

%, which is determined by %criteria ultimately based on %the maximization of the %overall power transfer to the UE. 
Our proposed techniques (called SLO and ALO) have been obtained by optimizing, at each point of the IRS, the power transfer of the most significant components of the signal spectrum, and by using the generalized Snell's law as the ground to calculate the phase shift along the IRS surface. An upper bound (UB) and narrowband-based (NB and NBF) approaches serve as benchmark. Comparison is also made against another technique from the literature, relying on eigenvalue decomposition (ED), which globally considers the IRS surface as a whole.

The performance of all these mentioned techniques are compared in terms of the transfer function they provide, as well as in terms of the average received power spectral density. We also considered a specific coefficient of variation for the overall frequency response, which measures the distortion of the received spectrum with respect to the transmitted one. The analysis, made for relative bandwidths ranging from $10\%$ up to $40\%$, shows a clear advantage of the proposed methods (SLO and ALO) especially when the IRS is small and when the transmitted signal spectrum is wideband, is composed of more than one sub-band and/or it is not flat.
Future work will delve deeper into these findings, testing the application of SLO and ALO to other system geometries and bigger IRSs, where signal self-interference among IRS elements has to be countered.

\bibliographystyle{IEEEtran}
\bibliography{References}

\end{document}